\documentclass[10pt]{article}
\usepackage{graphicx}
\usepackage{amsfonts}
\usepackage{amssymb}
\usepackage{amsmath}
\usepackage{multicol}

\begin{document}

\title{N-fold scattering series for Kolmogorov equation}
\author{Alexei Buzdin, Sergey Leble\\ \small Immanuele Kant Baltic Federal University,  Al. Nevsky st 41, Kaliningrad,  Russia } 

\maketitle

\begin{abstract}
We consider a formulation of initial problem for Kolmogorov equation
that corresponds a localized source of particles to be scattered by medium with given scattering amplitude density (scattering indicatrix). The multiple scattering expansion and amplitudes are introduced and the corresponding series solution of the equation is constructed. We investigate the multiple integral representation for the series terms, transform it into a form, convenient for estimations and prove  convergence of the series. An application to light beam scattering and LIDAR problem solution is outlined.
\end{abstract}

\section{Introduction}
There are lot of issues representing methods of description of particles scattering in inhomogeneous media.
One of the  general transfer function and its most developed double scattering example from \cite{Eloranta} directly relates to a LIDAR backward probing.

 Another, more close to classical Fresnel-Kirchoff diffraction theory, it is applied to microwave propagation/scattering \cite{Carlson}. Next, interpreting LIDAR data, even in the single-scattering approximation, it is quite  complicated task based on the extinction coefficient. The double-path atmospheric transmission is also studied. Such method is effective for both the wave and the quantum natures of light and it is free from limitations on the forms of scattering phase function, multiplicity of photon interactions with a medium \cite{AO}. Physically, this behavior suggests that the radial distribution of the forward scattered photons spreads with increasing time delay. Comparison of the results obtained for a model clouds are in a good coincidence with the results of the double scatter model of \cite{Eloranta}. Further development to a random media case with account of polarization phase of photons one founds for example in \cite{Jake}.

 In the realm of applicability of extinction coefficients and scattering indicatrix  some developments for one- and double scattering were proposed in \cite{kaul} \cite{KolUch} and the stochasticity importance  is explained in 
 \cite{Uch}.

An alternative approach based on Kolmogorov equations was developed by authors in \cite{Tomsk}. A generalized Cauchy problem is solved in a series form by iterations in terms of distributions, obtaining results by averaging procedure over a receiver.

 Earlier version of one of Kolmogorov equations (forward one) have the physical origin related to Brownian motion, which is often named now as FOKKER-PLANCK EQUATION \cite{Fok,Plank}. A  diffusion form of the equation
\begin{equation}\label{FP}
   f_t=- \sum_I[b^if]_{x_i}+\frac{1}{2}\sum_{i,j}[a_{ij}f]_{x_ix_j},
\end{equation}
relates to general Kolmogorov equations for one-parametric family of continuous linear operators
$\mathbf{P}(t),\; t>0$ in topological vector space that expresses semigroup property:
\begin{equation}\label{gen}
    \mathbf{P}(t+s)=\mathbf{P}(t)\mathbf{P}(s).
\end{equation}
Such way we restrict ourselves  to homogeneous Markovian stochastic processes, where $\mathbf{P}(t),\; t\geq 0$   the operator, that transforms probability distribution at zero time to one of the time  t ($\mathbf{P}(0)=\mathbf{1}$).

Differentiating \eqref{gen} with respect to t, one arrives at
\begin{equation}
 \frac{d\mathbf{P}(t)}{dt}=\mathbf{P}(t)\mathbf{Q},
\end{equation}
where
\begin{equation}
 \mathbf{Q} =\lim_{h\rightarrow 0}\frac{\mathbf{P}(h)-\mathbf{1}}{h}.
\end{equation}
If one goes down to homogeneous Markovian stochastic processes in $x,y\in\mathbb{R^n}$
 (Chapman-Kolmogorov \cite{Ch},) it
is applied to many problems of physics as well \cite{Pasc}. Let us mention a book of \cite{KolUch}, that reviews a theory of neutron scattering problems.

The Chapman-Kolmogorov equation for transition densities directly generalizes \eqref{gen}:
\begin{equation}\label{ChK}
   p(t+s,x,y)=\int_{\mathbb{R^n}}p(t,x,z)p(s,z,y)dz.
\end{equation}

In this paper we start (Sec. \ref{Pf}) with a formulation of problem for the Kolmogorov equations for random dynamic processes \cite{K1931}. The form of the equation to be studied is intimately close to the linearized Boltzmann one \cite{Tomsk} with its direct physical sense as balance of scattering gas particles in phase space,
which corresponds a localized source of particles to be scattered by medium with given scattering amplitude density. The multiple scattering amplitudes are introduced and the corresponding series solution of the equation is constructed (Sec. \ref{MS}). We investigate the multiple integral representation for the series terms, its estimations and
prove  convergence of the series. We also present explicit expressions for the first terms of the expansion and corresponding particles rates (Sec. \ref{two}).  An application to photons beam scattering is considered
\cite{BL1980,GL}. 

Most important aim which we would like to achieve is to investigate the iteration series
  convergence in some natural conditions (Sec. \ref{conv}). The upper estimation of the terms for a point receiver (Sec. \ref{rates} ) is achieved by division of integration domain in a space of specially constructed variables and give a possibility to estimate N-fold scattering contributions in terms of problem parameters.

\section{Problem formulation}\label{Pf}

\subsection{Initial problem}

Consider a kinetic equation in the  phase space $\{\vec r, \vec v\}\in \Gamma$
\begin{equation}\label{E:con_CNLSE_2_1}
 \partial_{t}f+\vec v\cdot\nabla f=-\sigma_t(z)
f-\int_{\Omega}\sigma(cos\gamma, z)fd\Omega',
\end{equation}
where $f(t,\vec r,\vec v)$ is proportional to probability density  function over $\{\vec r,\vec v\}\in \Gamma$, $\vec v=(v sin \theta cos\phi, v\sin\theta sin\phi, v cos \theta)$ and $\sigma_t(z),\sigma(cos\gamma, z) $ are   total and differential cross section densities per unit volume of a stratified (in z) medium correspondingly. The function $f(t,\vec r,\vec v)$ itself will be used as a distribution defined by action on the Schwartz space   $\psi(\vec r,\theta,\phi)\in\textbf{S}$, via continuous linear functional  $(f,\psi)\in R$. In spherical coordinates of incident $\theta,\phi$ and scattered $\theta',\phi'$ particles the scattering angle is characterized by
\begin{equation}\label{cosG}
   \cos\gamma = \cos\theta \cos\theta'+\sin\theta \sin\theta'\cos(\phi-\phi'),
\end{equation}
hence $d\Omega'=sin\theta'd \theta' d\phi'.$ We suppose that the scattering is elastic, $|\vec v|$ does not change while the scattering process occur,  we can choose units so that $|\vec v|=1$.
Initial condition for (\ref{E:con_CNLSE_2_1}) is represented by a distribution, this paper choice is restricted by 
\begin{equation}\label{ini}
  f(0,\vec r,\theta,\phi) =  V\delta(\vec r)\delta(\theta),
\end{equation}
with a constant $V$ as normalization factor. It means that we built a solution for the probability density as a weak limit (when $t \rightarrow 0$) to the $\delta$ - function at $t=0$.
The distribution $\delta(\theta)$ is chosen as
\begin{equation}\label{deltateta}
    (\delta(\theta),\psi(\vec r,\theta,\phi))=\int_0^{2\pi}\psi(\vec r,0,\phi)d\phi,
\end{equation}
$\psi\in\textbf{S}$, and, in a conventional mode,  $\delta(\vec r)=\delta(x)\delta(y)\delta(z)$.

The distribution $f$ is defined in a (Schwartz) space which will be used in direct applications via the integral by space variables which enter the solution as parameters, used as a receiver geometry description.  
So, we are interested in
\begin{equation}\label{JD}
   J(\Delta , t)=\int_{\theta_0}^{\theta_1}(f(t,\vec r,\theta,\phi),\psi)d\theta  
\end{equation}
The applications relate to observations (measurements) of the result of averaging procedure, defined by \eqref{JD}.
The expression (\ref{JD}) defines number of particles within a finite domain ($\Delta$) of a measurement apparatus and having velocity direction between $\theta_0$ and $\theta_1$ restricted by aperture related to the apparatus window direction. The action of the distribution $f$ on $\textbf{S}$ in the case of piecewise continuous functions implies the integration with respect to $\phi,x,y,z$. 

An example of conventional backward (LIDAR) receiver posed in a vicinity of the origin,  the function  $\psi$ was chosen  nonzero (e.g. equal to 1)        if

$$
  \begin{array}{c}
    |x|\leq\Delta x, \\
    |y|\leq\Delta y,\\
    |z|\leq\Delta t|cos\theta |,
  \end{array}
$$
where 
$\Delta t $ is the receiver reaction time and the aperture angles lie between $\pi$ and $\pi-\theta_0$.

We would use also the point receiver that is convenient when the receiver is small compared with the solution inhomogeneity scale.
 \begin{equation}\label{Jlim}
    J_p( t)=\lim_{\Delta\rightarrow 0}\frac{J(\Delta,t)}{\Delta}.
  \end{equation}

   We will also use such function for estimation of N-fold scattering contribution and for investigation
of the series convergence in the Sec. \ref{Con}.

 Having in mind applications to electromagnetic wave scattering problem, the function $f$ is interpreted as either a density number of  a photons or probability density in $\Gamma$ space.

\subsection{Number of particles rate}
\label{NPR}

Generally \cite{Tomsk}, the probabilistic  interpretation of the distribution function $f$ in the phase space gives the number of particles in a small volume $\Delta x \Delta y\Delta z$ as
$$
\int_0^{2\pi}\int_{\pi-\theta_0}^{\pi}\int_x^{x+\Delta x}\int_y^{y+\Delta y}\int_z^{z+\Delta z} f dxdydz \sin\theta d\theta d\phi.
$$
Whereas in the LIDAR problem the receiver is placed at the origin \cite{Tomsk}, generally it may be placed at a certain position outside the scatterers  domain.

   By its direct sense, the particles rate is proportional to
the number of particles (e.g. photons) per unit time and volume, that is given by general relation
\begin{equation}\label{Intensity}
I(t) = \lim\limits_{\Delta t \rightarrow 0} \frac{1}{\Delta t} \int_{0}^{\alpha} (f(t,0,0,z_0,\theta,\phi),\psi(x,y,z,\theta,\phi))\sin\theta   d\theta.
\end{equation} 
The choice of the function $\psi$ can be realized by concrete physical reasons. 

   Our  particular aim is the evaluation of number of particles per unit time which enter the round area of radius $\rho_0$ laying in the plane $z = z_0$ (receiver) with center in the origin and having velocity vectors inclined to z-axis within the angle interval $\theta \in [0, \theta_0]$. Here an aperture angle $\theta_0$,  restricts possible velocities of particles directions. 
In the exemplary case we take here, the receiver has cylindrical symmetry and for the initial direction along $z$, the function does not depend on $\theta, \phi $, so it is chosen zero outside the receiver, and  $\psi(x,y,z) = 1$ for internal points of the domain $x^2 + y^2 \leq \rho_0^2$, \qquad $z_0 \leq z \leq z_0 + \Delta t |\cos \theta|$ and zero outside, being $z_0$ the coordinate and $|z_0|$ of the distance between the source of the pulse and the receiver.

\begin{equation}\label{Intensrho}
I(t) = \lim\limits_{\Delta t \rightarrow 0} \frac{1}{\Delta t} \int_0^{\rho_0}  \int_{0}^{\alpha}\int_0^{2\pi}(f(t,0,0,z_0,\theta,\phi),\psi(x,y,z,\theta,\phi))\sin\theta d\phi d\theta.
\end{equation} 
 
 For a point receiver we will go to the limit
 \begin{equation}\label{Intensrhop}
I_p(t) = \lim\limits_{\Delta t \rightarrow 0} \frac{1}{\Delta t}\lim\limits_{\rho_0 \rightarrow 0} \int_0^{\rho_0}  \int_{0}^{\alpha}\int_0^{2\pi}(f(t,0,0,z_0,\theta,\phi),\psi(x,y,z,\theta,\phi))\sin\theta d\phi d\theta.
\end{equation} 
  Such geometry was used in \cite{GL} for X-rays scattering problem application.
  
\section{Method of solution}\label{MS}

A solution is searched as a N-fold scattering expansion
\begin{equation}\label{U}
    f=f_0+f_1+f_2+... .
\end{equation}
We choose for $f_0$ the equation
\begin{equation}\label{Ueq}
 Lf_0=\frac{\partial f_0}{\partial t}+cos \theta\frac{\partial f_0}{\partial z}+sin\theta cos\phi\frac{\partial f_0}{\partial x}+sin\theta sin\phi\frac{\partial f_0}{\partial y}=-\sigma_t f_0,
\end{equation}
and initial condition as the distribution \eqref{ini},\eqref{deltateta}  where $V$ is chosen from normalization as the probability density.
\begin{equation}\label{ini0}
  f_0(0,\vec r,\theta,\phi) =  \frac{1}{2\pi}\delta(\vec r)\delta(\theta).
\end{equation}

For $n>1$, the expansion coefficients are defined by
\begin{equation}\label{fn}
    L f_{n+1}=-\sigma_t f_n+\int_0^{2\pi}\int_0^{\pi}\sigma(cos\gamma, z) f_n sin\theta'd\theta'd\phi'.
\end{equation}
with zero initial conditions for $n>0$
\begin{equation}\label{in}
    f_n|_{t=0}=0.
\end{equation}

Let us change variables in \eqref{Ueq} 
\begin{subequations}
\begin{eqnarray}
x'=x-t  \sin\theta \cos \phi ,\\
y'=y-t\sin\theta sin \phi,\\
z'=z-t\cos\theta.
\end{eqnarray}
\end{subequations}
The equation \eqref{Ueq} is transformed as
\begin{equation}\label{Ueqtr}
   \frac{\partial f_0}{\partial t}=-\sigma_t f_0
\end{equation}
which is integrated as
\begin{equation}\label{U0}
    f_0=G(\vec r,\vec v )\exp[-\int_0^t\sigma_{t}(z')d\tau].
\end{equation}
The function $G$ is found from initial conditions \eqref{ini0}:
\begin{equation}\label{ini0i}
   f_0(0,\vec r,\theta,\phi) =  V\delta(x)\delta(y)\delta(z)\delta(\theta).
\end{equation}
It gives
\begin{equation}\label{incond}
   G(\vec r,\vec v)=  V\delta(z-t)\delta(x)\delta(y)\delta(\theta).
\end{equation}
Let us denote a  function $E$ via  
\begin{equation}\label{E}
    E(t,z,\theta) =\exp[-\int_0^t\sigma_{t}(z-\tau cos\theta)]d\tau,
\end{equation}
hence, picking up (\ref{U0},\ref{incond}) and \eqref{E} for the $f_{0}$ one obtains 
\begin{equation}\label{U01}
    f_0(t,\vec r,\theta,\phi)=V\delta(z-t)\delta(x)\delta(y)\delta(\theta)E(t,z,\theta).
\end{equation}

Fo the n-fold iteration $f_{n+1}$, one has
\begin{equation}\label{fn+1}
\begin{array}{c}
f_{n+1}(x,y,z,\theta,\phi)=\\
   \int_0^tE(\tau,z,\theta)\int_0^{\pi}\int_0^{2\pi}\sigma(cos\gamma,z-\tau cos\theta)\\ f_{n}(t-\tau,x-\tau sin  \theta cos \phi, y -\tau sin  \theta sin \phi, z- \tau cos\theta, \theta',\phi')sin\theta' d\theta' d\phi' d\tau.
\end{array}
\end{equation}
This expression defines the  recurrence operator
\begin{equation}\label{Kn}
	f_{n+1}=K_{n}f_{n},
\end{equation}
which form decides about properties of approximate solutions and  convergence of the multiple scattering series \eqref{U}.
%In particular, for $f_1$ and $f_2$ we obtain the following expressions.

For the first iteration  $f_{1}$ one obtains:
\begin{equation}\label{U1}
\begin{array}{c}
  f_{1}=V\int_{0}^t E(\tau,z,\theta)E(t-\tau,z-\tau cos\theta,0)  \\ \sigma(cos\theta,z-\tau cos\theta)
  \delta(x-\tau cos\theta-(t-\tau))  \delta(y-\tau sin\theta sin\phi)\delta(z-\tau sin\theta cos\phi )d\tau .
  \end{array}
\end{equation}

 Similar expressions are obtained and interpreted in a case of N-fold scattering  terms, the two-fold one  is presented  in the following sections (see also \cite{BL1980}). 

 \section{Examples of particles rates estimation.} 
 \label{rates}
 
To explain the approach based on transition to new variables to be convenient for the N-fold contributions in the expansion \eqref{U} we would demonstrate its form on minimal number of iterations.  

\subsection{An one-fold solution for a point receiver}

The  integrand  in \eqref{U1}   is considered as distributions on Schwartz space $\mathbf{S}$ of functions $x,y,z$ which depend on $\phi,\theta, \tau$ as parameters. For example, $f_{1}$ acts on an element  $\psi(\vec r) \in \mathbf{S}$ as
\begin{equation}\label{act}
\begin{array}{c}
 ( f_{1}(\vec r,\theta,\phi,t),\psi)=
  \\
V\int_0^{2\pi}\int_0^tE(\tau,\tau\cos\theta+t-\tau,\theta)E(t-\tau,t-\tau,0)\sigma (\cos\theta,\tau\cos\theta+t-\tau) 
\\ 
\psi(\tau\sin\theta cos\phi,\tau\sin\theta sin\phi,\tau\cos\theta+t-\tau)d\tau d\phi,
\end{array}
 \end{equation}
 where $\psi$ describes a receiver.  For example it is determined as in Sec. \eqref{NPR}, or  $\psi(\vec r)=1$ at 
  $$x^2 + y^2 \leq \rho_0^2, \qquad z_0 \leq z \leq z_0 + \Delta t |\cos \theta|$$
   and zero outside, being $z_0$ the boundary coordinate of the cylindrical receiver.
 
From the definition \eqref{JD} for the backward scattering and aperture angle $\theta_0$ we have
 \begin{equation}\label{Jpsi1}
    J_1(t)=\int_{\theta_0}^{\pi} (f_1,\psi)sin\theta d\theta.
  \end{equation}

 Therefore, we get for  \eqref{U1} going to the intensity \eqref{Intensity} for a point receiver that we do as the limit:
\begin{equation}\label{f1I11}
\begin{array}{c}
	I_{1p} (t,0,0,z_0)= \\ -  \lim\limits_{\Delta t \rightarrow 0} \lim\limits_{\rho_0 \rightarrow 0}\int_{z_0}^{z_0+\cos\theta_0\Delta t}\int_0^{2\pi}\int_0^{\rho_0}\int_0^{\alpha}\int_0^t  E(\tau,z_0,\theta)
	\int\sigma_{scat}(\cos \theta \cos \theta'  , z-\tau\cos\theta)\\f_0 (t-\tau, x-\tau\sin\theta\cos\phi,y-\tau\sin\theta\sin\phi,z-\tau\cos\theta,\theta)\sin\theta'd\theta'd\tau d\theta d\rho d\phi dz,
\end{array}
\end{equation}
the normalizing factor is chosen as $V=1/2\pi$, integrating by $\phi$, taking the cylindrical symmetry into account. Evaluation of 1-fold scattering by \eqref{act} 
\begin{equation}\label{f1I12}
\begin{array}{c}
I_{1p}(t) = \\ - \lim\limits_{\Delta t \rightarrow 0} \lim\limits_{\rho_0 \rightarrow 0} \frac{1}{\Delta t} \int_0^{\rho_0}\int_0^{\alpha} \int_0^t\int_{z_0}^{z_0+\cos\theta_0\Delta t} E(\tau,z,\theta_0)  E(t-\tau,z-\tau\cos\theta_0,\theta_0)\\
\int	\sigma_{scat}(\cos \theta_0 \cos \theta'  , z-\tau\cos\theta_0)  \psi(\tau\sin\theta_0\cos\phi ,\tau\sin\theta_0\sin\phi 
	,\tau\cos\theta_0)\sin\theta'd\theta'd\tau d\phi d\rho dz.
\end{array}
\end{equation}
 The nonzero $I_1$ values are obtained in conditions that are explained by means of the following description.  
	 The area of integration lies between the horizontal lines $z=z_0,z_0+\Delta z$ and inclined lines $z=\cos\theta_0+a$ and $z=\cos\theta_0+b$, where a,b are boundaries of domain, filled with scatterers under consideration: e.g.  a "cloud" in atmosphere \cite{KS1} or metal plate in X-rays case \cite{GL}. The vertical green line marks a pulse arrival time $z=t$ (see also the Fig.1 in \cite{GL}).

In the case of fixed angle $\theta_0$,  
$z' = \tau (\cos \theta_0 - 1)+ t $, 
$
\tau=(\cos \theta_0 - 1)^{-1}(z'-t),
$
hence the argument of the $\delta$ - function
is
$
z_0+t-\tau-z' =z_0+t-(\cos \theta_0 - 1)^{-1}(z'-t)-z'=z_0-bt+az'=a(z'-\frac{b}{a})t+z_0/a, 
$
where
$a=-1-(\cos \theta_0 - 1)^{-1}=\frac{\cos \theta_0}{1-\cos \theta_0},\,b=1+(\cos \theta_0 - 1)^{-1}.$

The second argument of the scattering amplitude $\sigma_{s}$ is therefore $z_0-\tau\cos\theta_0=z_0-(\cos \theta_0 - 1)^{-1}(z'-t)\cos\theta_0.$

 The result of 1-fold scattering  for zero angle for the point receiver is almost trivial from geometrical point of view, the arriving pulse is infinitely short. The expression for intensity   contains natural spherical divergence, exponential decay due to absorption and forward scattering in a level inside the layer. Some details may be found in the papers \cite{GL}.   

and \eqref{Jlim}
 \begin{equation}\label{Jlim1}
    J_{1}^p(t)=lim_{\Delta_i\rightarrow 0}\frac{J_1(t)}{\prod\Delta_i}, i=1,2,3.
  \end{equation}
one arrives at
\begin{equation}\label{Jlim11}
    J_{1}^p(t)=\lim_{\Delta_i\rightarrow 0}\frac{1}{\prod\Delta_i}\int_{\theta_0}^{\pi}\int_0^{2\pi}(f_1,\psi)sin\theta d\phi d\theta, i=1,2,3.
  \end{equation}
where the vector $\psi$ have nonzero components if

$$
  \begin{array}{c}
    |x|\leq\Delta x=\Delta_1, \\
    |y|\leq\Delta y=\Delta_3,\\
    |z|\leq\Delta t|cos\theta |=\Delta_3.
  \end{array}
$$
and $0$ outside the domain. Plugging $f_1$  from \eqref{U1} yields:
\begin{equation}\label{Jlim12}
\begin{array}{c}
    J_{1}^p(t)=\lim_{\Delta_i\rightarrow 0}\frac{1}{\prod\Delta_i}\int_{\theta_0}^{\pi}\int_0^{2\pi}\int_0^tE(\tau,\tau\cos\theta+t-\tau,\theta)E(t-\tau,t-\tau,0)\cdot\\
 \sigma (\cos\theta,\tau\cos\theta+t-\tau)\psi(\tau\sin\theta cos\phi,\tau \sin\theta \sin\phi,\tau\cos\theta+t-\tau)d\tau\sin\theta d\phi d\theta.
\end{array}
  \end{equation}

\subsection{Alternative variables of integration. One- and two-fold solutions for a point receiver}
\label{two}

Let us change the variables of integration, having in mind more convenient (compared to \cite{BL1980}) description of the integration domain and limiting procedure
\begin{equation}\label{var}
     \begin{array}{c}
       x=\tau\sin\theta \cos\phi, \\
      y=\tau\sin\theta \sin\phi,  \\
      x=\tau\cos\theta+t-\tau.
     \end{array}
\end{equation}
The inverse ones are found as
\begin{equation}\label{invvar}
     \begin{array}{c}
        \tau=\frac{x^2+y^2+z^2-2zt+t^2}{2(t-z)}, \\
       \cos\phi=\frac{x}{\sqrt{x^2+y^2}}, \\
       \cos\theta=\frac{x^2+y^2+z^2-2zt+t^2-2(t-z)^2}{x^2+y^2+z^2-2tz+t^2),}
     \end{array}
\end{equation}
with the Jacobian:
\begin{equation}\label{J}
    J\sin\theta=\frac{2}{x^2+y^2+z^2-2zt+t^2}.
\end{equation}
Let us define the integration intervals by means of
\begin{equation}\label{inter}
  \begin{array}{c}
    0\leq\tau\leq t,\\
     0\leq\phi\leq 2\pi, \\
 \pi-\theta_0\leq\theta\leq \pi,\\
     0\leq\tau \cos\theta +t-\tau\leq \Delta t, \\
    |\tau\sin \theta \cos\phi|\leq \Delta x, \\
   |\tau\sin \theta \sin\phi|\leq \Delta y,  \\
   t-\tau +\tau\cos\theta \geq 0,
  \end{array}
\end{equation}
then $|x|\leq \Delta x$, $|y|\leq \Delta y$, $0\leq z\leq \Delta t$.

The explicit expression  for the integral $J_1$, corresponding to \eqref{Jlim12},
in new variables  takes the form
\begin{equation}\label{JJ}
    J_1=\lim \frac{1}{2\Delta x\Delta y\Delta z}\int_{-\Delta x}^{\Delta x}\int_{-\Delta y}^{\Delta y}\int_{-\Delta t}^{0}   \frac{E'E''\sigma }{x^2+y^2+z^2-2zt+t^2} dxdydz,
\end{equation}
where $$E'=E (\tau,z,\theta),\quad E''=E (t-\tau,z-\tau \cos\theta,0),$$
is defined by (\ref{E}).
In arguments of $E',E'',\sigma$ the old variables $\tau,\theta$ are expressed in new ones by (\ref{invvar}).

Doing the limiting transition $\Delta x\rightarrow 0,\Delta y\rightarrow 0, \Delta z\rightarrow 0, \Delta t\rightarrow 0$ for a point receiver, we obtain the simple formula for back scattering  
\begin{equation}\label{I1}
    I_1^p(t)=    \frac{2}{t^2}E'(\frac{t}{2},0,\pi)E''(\frac{t}{2},\frac{t}{2},0)\sigma(-1,\frac{t}{2}). 
    \end{equation}
It reproduces the one-fold LIDAR formula for a small receiver in convenient form, see the recent \cite{KS1}.

From \eqref{fn+1} similar to \eqref{U1} using the correspondingly modified transformations (\ref{var},\ref{invvar}) one arrives at the distribution term $f_2$:

\begin{equation}\label{U2}
\begin{array}{c}
f_{2}=
\int_0^t E(\tau_2,z,\theta)\int_0^{2\pi}\int_0^{\pi}\sigma(cos\gamma,z-\tau_2 cos\theta)\\
\int_0^{t-\tau_2} E(\tau,z-\tau_2 cos\theta,\theta')E(t-\tau_2-\tau,t-\tau-\tau_2,0)\sigma(cos\theta' ,t-\tau-\tau_2)\\\delta(x-\tau_2 sin\theta cos\phi-\tau sin\theta' cos\phi')\delta(y-\tau_2 sin\theta sin\phi-\tau sin\theta' sin\phi')\\
\delta(z-\tau_2 cos\theta-\tau cos\theta'-(t-\tau_2-\tau))d\tau sin\theta'd\theta'd\phi'd\tau_2
  \end{array}
\end{equation}
In new variables \eqref{var} the 
integral $J_2$ takes the form
\begin{equation}\label{JJ}
\begin{array}{c}
    J_2=\lim_{\Delta V\rightarrow 0} \frac{1}{2\Delta x\Delta y\Delta z}\int_{-\Delta x}^{\Delta x}\int_{-\Delta y}^{\Delta y}\int_{0}^{\Delta t}\int_{0}^{t/2}\int_{0}^{2\pi}\int_{\pi-\theta_0}^{\pi} \\
     \frac{E^* \sigma(\cos \gamma,  \tau_1( \cos \theta_1-1)+t-\tau_2) \sigma(\cos \theta_1,t-  \tau_1-\tau_2)}{(\tau_2sin\theta_2cos\phi_2-x)^2+(\tau_2sin\theta_2sin\phi_2-y)^2+(t-(1-cos\theta_2)\tau_2-z)^2} 
   sin\theta_2d\theta_2d\phi_2d\tau_2 dzdxdy,
       \end{array}
\end{equation}
where
\begin{equation}\label{E*}
	E^*=E(\tau,z-\tau_2 cos\theta,\theta')E(t-\tau_2-\tau,t-\tau-\tau_2,0).
\end{equation}

 Approximate expressions for the main term contribution in the two-fold scattering are presented in \cite{BL1980} as double integral and compared to \cite{kaul,KKK}.
 
 \section{Multiple scattering series solution} 
 \label{conv}
 
\subsection{n-fold solution for a point receiver}
The algorithm prescribed by the recurrence relation \eqref{Kn} yields
\begin{equation}\label{JN1}
    J_n=\lim_{\Delta_i\rightarrow 0} \frac{1}{\prod_i \Delta_i}\int_{\pi-\theta_0}^{\pi}\int_0^{2\pi}(f_{n},\psi) sin\theta d\theta d\phi,
\end{equation}
the nonzero contribution in the integrand of \eqref{JN1} is defined by  
\begin{equation}\label{JN}
\begin{array}{c}
    J_n= \\ \lim_{\Delta_i\rightarrow 0} \frac{1}{\prod \Delta_i}\int_{-\Delta x}^{\Delta x}\int_{-\Delta  y}^{\Delta y}\int_{-\Delta t}^{0}\int_{\pi-\theta_0}^{\pi}\int_0^{2\pi}\int_0^t...\int_0^{t -\sum_{i=1}^{n-1}\tau_i}E^*\sigma(\cos\gamma,z-\tau_n \cos\theta)...\\ \sigma(cos \theta_1,z-\sum_{i=1}^n\tau_i\cos\theta_i) \\ \psi (\sum_{i=1}^{n}\tau_i\sin\theta_i\cos\phi_i,\sum_{i=1}^{n}\tau_i\sin\theta_i\sin\phi_i,\sum_{i=1}^{n}\tau_i\cos\theta_i+t-\sum_{i=1}^{n}\tau_i)\\
    \prod_{i=1}^{n}(d\phi_id\tau_id\theta_i)dV,
    \end{array}
\end{equation}
where $E^*$ is a direct generalization of the exponential functions product \eqref{E*} that characterize the attenuation   along the whole scattering path of a particle.
 
\subsection{Estimation of n-fold contribution for a point receiver}

The analysis of  the first terms (\ref{U1},\ref{U2}) of the series \eqref{U} gives an idea to write general expressions and perform  estimations. The transformation \eqref{var} now looks as 
\begin{equation}\label{Tr}
\begin{array}{c}
     x=\sum_{i=1}^n\tau_i\sin\theta_i\cos\phi_i\\  
     y= \sum_{i=1}^n\tau_i\sin\theta_i\sin\phi_i\\
    z=\sum_{i=1}^n\tau_i\cos\theta_i+t-\sum\tau_i,
    \end{array}
    \end{equation}
    its inverse for $\tau_1,\cos \theta_1$ and $\cos\phi_1$ and Jacobian look similar to (\ref{var},\ref{invvar},\ref{J}). 
For the $n$ times scattering  contribution, after transition to the limit, one has
\begin{equation}\label{JN}\begin{array}{c}
    J_n= 2 \int_{D} E^*\sigma(cos\gamma,\sum_{i=1}^{n-1}\tau_i\cos\theta_i t-\sum\tau_i) \bullet...\\ \bullet\frac{\sigma (\cos\theta_1,t-\sum_{i=1}^{n}\tau_i)\prod_{i=2}^{n}d\phi_id\tau_id\theta_i}{(t-\sum_2^n\tau_i(1-cos\theta_i))^2+ (t-\sum_2^n\tau_i cos\theta_i sin\phi_i)^2+(\sum_2^n\tau_i cos \theta_i cos\phi_i)^2},
    \end{array}
\end{equation}
where the integration domain $D$ is specified by the inequalities
\begin{equation}\label{JN2}
\begin{array}{c}
     0<\theta_2<\pi,...,\pi-\theta_0<\theta_n<\pi,\\
     0<\phi_2<2\pi,..., 0<\phi_n<2\pi,\\
     (t-\sum_2^n\tau_i)^2\geq \sum_2^n\tau_i^2+\sum_{i\neq k}\tau_i\tau_k cos\gamma_{ik},\\
     \tau_i \geq 0,\\
      \gamma_{ik}=cos\theta_i cos\theta_l+sin\theta_i sin\theta_k cos(\phi_i-\phi_k).
    \end{array}
\end{equation}
Let us divide the integration domain $D$ to subdomains $D_1,D_2$ so that for a positive $\epsilon_n$ holds
\begin{equation}\label{D12} 
(t-\sum_2^n\tau_i(1-cos\theta_i))^2+ (\sum_2^n\tau_i \cos\theta_i sin\phi_i)^2+(\sum_2^n\tau_i cos \theta_i cos\phi_i)^2
\begin{array}{c}
 \leq   \epsilon_n \sim D_1,\\
   \geq  \epsilon_n \sim D_2 .
    \end{array}
\end{equation}
Denote the correspondent integrals as $J_n^1, J_n^2.$

\subsection{Multiple scattering series convergence theorem}\label{Con}

Let the Chebyshev  norm of $\sigma$   is denoted as $||\sigma||=\max_{\cos\gamma\in[-1,1],z\in [z_1,z_2]}\sigma(\cos\gamma,z)$, the scatterers now are within the interval $[z_1,z_2]$.
After estimations in integrands we arrive at
\begin{equation}\label{Jn12}
\begin{array}{c}
J_n^1 \leq 2^{n-1}||\sigma||^n\exp[-\sigma_{min}t]\frac{(n-1)^3t^{n-2}\epsilon_n(4\pi)^{n-1}}{(t-\epsilon_n)^2}\\
J_n^2\leq 2^{n}||\sigma||^n\exp[-\sigma_{min}t]\frac{(4\pi t)^{n-1}}{\epsilon_n^2(n-1)!}\theta_0^2.
	   \end{array}
\end{equation}
Choosing next 
\begin{equation}
	\epsilon_n= \frac{ \theta_0^{2/3}t}{2((n-1)!)^{1/3}(n-1)},
\end{equation}
and unifying \eqref{Jn12} we obtain the Resulting estimation
\begin{equation}\label{JNE}
\begin{array}{c}
    J_n  \leq \\
    2^{3n-1}  ||\sigma||^n\exp{[-\sigma_{min}t]}\pi^{n-1}t^{n-3}\frac{(n-1)^2}{((n-1)!)^{1/3}}\theta_0^{2/3}.
    \end{array}
\end{equation}
The series generated by r.h.s. of the inequality \eqref{JNE} converge because
\begin{equation}
	\lim_{n\rightarrow\infty} \sqrt[n]{2^{3n-1}  ||\sigma||^n\exp{[-\sigma_{min}t]}\pi^{n-1}t^{n-3}\frac{(n-1)^2}{((n-1)!)^{1/3}}}\rightarrow 0.
\end{equation}
Therefore radius of convergence is infinity.

If $8\pi||\sigma||t<1,$ the following estimate for error is following
\begin{equation}
\begin{array}{c}
	\sum_1^n I_k(t)\leq  4||\sigma||e^{-\sigma_{min}t} \theta_0^{2/3} t^2\sum_{n=k}^{\infty}(8\pi||\sigma||t)^{n-1}=\\
	\frac{||\sigma||e^{-\sigma_{min}}t \theta_0^{2/3}}{1-8\pi||\sigma||t}.
	  \end{array}
\end{equation}
Having such estimation one can decide what number of N-fold contributions should be taken into account for a given error via approximate estimation for $\sigma$.

\section{Conclusion}
We study a construction and convergence of iterative series  solution of Kolmogorov equation  which would be
useful to approve and estimate a contribution of N-times scattering into the complete response.  There is a possibility to enhance
the method to third or higher order to improve evaluation of particles scattering in a medium of higher density. In the manuscript we use the "physical" language of particles as alternative to semi-analytical and Monte-Carlo approaches  \cite{CZK} but the result has obvious general mathematical sense for arbitrary events related to the theory of probability. Generalizations for polarization account \cite{kaul} are also evident and will be published elsewhere. 

\section{Acknowledgment}
 The work is supported by Ministry of Education and Science of the Russian Federation (Contracts No N GZ 3.1127.2014K)"

\end{document}